\documentclass[graybox]{svmult}

\usepackage{type1cm}        
\usepackage{makeidx}         
\usepackage{graphicx}        
                             
\usepackage{multicol}        
\usepackage[bottom]{footmisc}
\usepackage{amsmath,upref}

\usepackage{newtxtext}       %
\usepackage{newtxmath}       

\usepackage{ulem}

\usepackage{cite}
\usepackage{relsize}

\usepackage{labelfig}

\allowdisplaybreaks

\makeindex             

\DeclareMathAlphabet{\mathpzc}{OT1}{pzc}{m}{it}

\renewcommand{\E}{\mathrm{e}\kern0.2pt}%
\renewcommand{\D}{\mathrm{d}\kern0.2pt}%
\newcommand{\ann}{\mathsmaller{\boldsymbol\circ}}
\newcommand{\annB}{\mathsmaller{\boldsymbol\circ}}

\begin{document}

\title*{Sloshing in vertical cylinders with circular walls: \\ the effect of radial
baffles}
\author{Nikolay Kuznetsov, Oleg Motygin}
\institute{N. Kuznetsov \at
              Laboratory for Mathematical Modelling of Wave Phenomena, 
              Institute for Problems in Mechanical Engineering, 
              Russian Academy of Sciences, 
              V.O., Bol'shoy pr.\ 61, St.\,Petersburg 199178, Russian Federation
              \email{nikolay.g.kuznetsov@gmail.com}
\and O. Motygin \at
              Laboratory for Mathematical Modelling of Wave Phenomena, 
              Institute for Problems in Mechanical Engineering, 
              Russian Academy of Sciences, 
              V.O., Bol'shoy pr.\ 61, St.\,Petersburg 199178, Russian Federation
              \email{mov@ipme.ru, o.v.motygin@gmail.com}}
%
%

\maketitle


\abstract{The behaviour of sloshing eigenvalues and eigenfunctions is studied for vertical
cylindrical containers that have circular walls and constant (possibly infinite) depth. The effect of
breaking the axial symmetry due to the presence of radial baffles is analysed. It
occurs that the lowest eigenvalues are substantially smaller for containers with
baffles going throughout the depth; moreover, all eigenvalues are simple in this
case. On the other hand, the lowest eigenvalue has multiplicity two in the absence
of baffle. It is shown how these properties affect the location of maxima and minima
of the free surface elevation and the location of its nodes.
}


\section{Introduction}
\label{intro}

The sloshing of a fluid in various partially filled containers is a topic of great
interest to engineers, physicists and mathematicians. It has received extensive
study; see, for example, the monographs \cite{FT} and \cite{I} (the second one has
more than 100 pages of references). A historical review going back to the 18th
century can be found in \cite{FK}, whereas the comprehensive book \cite{KK} presents
an advanced mathematical approach to the problem based on spectral theory of
operators in a Hilbert space. It is also worth mentioning that the 2012 Ig~Nobel
Prize for Fluid Dynamics was awarded to Krechetnikov and Mayer for their work
\cite{KM2012} concerning the dynamics of sloshing when walking with a mug of coffee.

The standard approach to sloshing is based on the linear water wave theory (see
again the books \cite{FT} and \cite{I}). In its framework, one seeks sloshing modes
and frequencies using eigenfunctions and eigenvalues, respectively, of a mixed
Steklov problem for the Laplace equation, and so a spectral parameter, say $\nu$,
appears in a boundary condition, whereas the radian frequency of fluid's
oscillations is proportional to the square root of $\nu$.

The question of suppression of sloshing by means of baffles (it is of importance for
practice) goes back to the 1960s. The corresponding experimental and theoretical
results were the topic of numerous NASA Technical Reports cited in \cite{I} (in
particular, a lot of data is summarised in \cite{A}). In many cases, it was found
that the effect of a baffle is to lower sloshing frequencies, especially, the lowest
one. Therefore, a lot of research has been carried out in this field since the
1960s. In particular, some quite sophisticated techniques were developed for
numerical evaluation of this effect during the past 30 years; here we mention a few
methods that look the most interesting. In \cite{EM}, eigenfunction expansions were
applied on either side of a vertical baffle in a rectangular tank. By matching these
expansions one obtains an integral equation on the interval below or above the
baffle. The approach developed in \cite{G}, uses fundamental solutions of the
problem for reducing it to an integral equation in the case of a vertical circular
container with a horizontal annular baffle. The method of successive conformal
mappings leading to standard truncated matrix eigenvalue problems which are then
solved numerically was used in \cite{HM}, where several different baffles in a
horizontal circular cylinder were considered.

Despite abundant numerical and analytical results (see e.g. \cite{Choudhary2017} and references therein), the mechanism of the baffle effect is far from
being completely understood. The aim of this note is to show the crucial role of
breaking the axial symmetry in this phenomenon. For this purpose we compare
properties of explicit solutions obtained by separation of variables for two pairs
of vertical cylindrical containers.

First, we consider sloshing frequencies and the corresponding modes that describe
free oscillations of a fluid in a circular container of constant depth which has or
has not the following radial baffle. It is located between the cylinder's axis and
the vertical wall and goes throughout the container's depth. (According to Fox and
Kuttler \cite{FK}, it was Lord Rayleigh \cite{R} who completed the solution of the
problem without baffle, studies of which were initiated by Ostrogradsky and Poisson
in the 1820s.) The second pair of containers includes the vertical annular cylinder
without and with the radial baffle that connects the cylinder's walls orthogonally
to them and goes throughout the container's depth. To the author's knowledge, no
comparison of the described solutions was ever published. Thus, a gap concerning the
mechanism of the hydrodynamic effect due to radial baffles is filled in at least
partially.

\section{Statement of the problem}

The general formulation of the three-dimensional sloshing problem is as follows. A
fluid domain, say $W$, is bounded from above by a free surface and from below
(and/or laterally) by the wetted rigid part of the container's boundary. The
horizontal mean position of the free surface is a bounded two-dimensional domain,
say $F$, that can be multiply connected, whereas $\partial W \setminus F$ is a
piecewise-smooth surface whose boundary coincides with $\partial F$. Let Cartesian
coordinates $(x,y,z)$ be chosen so that the $z$-axis points upwards, the domain $F$
belongs to the plane $z = 0$ and the rest part of $\partial W$ lies in the
half-space $z < 0$.

The usual hydrodynamic assumptions are as follows: the surface tension is neglected
on the free surface; the fluid is inviscid, incompressible and heavy; its motion is
irrotational and of small amplitude. Then sloshing modes and frequencies are sought
using eigenfunctions and eigenvalues, respectively, of the following mixed Steklov
problem (see \cite{FT,I,FK,KK}):
\begin{align}
& \nabla^2 \phi = 0 \ \ \mbox{in} \ W , \label{slosh1} \\ 
& \phi_z = \nu \phi \ \
\mbox{on} \ F , \label{slosh2} \\ 
& \frac{\partial \phi}{\partial n} = 0 \ \
\mbox{on} \ \partial W \setminus \overline F, \label{slosh3} \\ 
& \int_F \phi \,\D x \D
y = 0 .
\label{slosh4}
\end{align}
Here $\nu$ is the spectral parameter and the last condition is imposed to exclude the
eigenfunction identically equal to a non-zero constant that corresponds to the zero
eigenvalue existing when the problem includes only relations
(\ref{slosh1})--(\ref{slosh3}).

In terms of $(\nu, \phi)$ found from problem (\ref{slosh1})--(\ref{slosh4}), the
velocity field of free oscillations of the fluid occupying $W$ is given by, say $\cos
(\omega t + \alpha) \nabla \phi (x,y,z)$, where $\alpha$ is a certain constant, $t$
stands for the time variable and $\omega = \sqrt{\nu g}$ is the radian frequency of
oscillations (as usual, $g$ denotes the acceleration due to gravity). Furthermore,
the elevation of the free surface is proportional to $\sin (\omega t + \alpha) \,
\phi (x,y,0)$.

It is known that problem (\ref{slosh1})--(\ref{slosh4}) has a sequence of
eigenvalues (see, for example, \cite{KK} and \cite{M}):
\[ 0 < \nu_1 \leq \nu_2 \leq \ldots \nu_n \leq \ldots , \quad \nu_n \to \infty ,
\]
to each of which a single eigenfunction $\phi_n$ corresponds, but if the
multiplicity of an eigenvalue is greater than one (however, it always is finite),
then this eigenvalue is repeated in the sequence as many times as the multiplicity
is. Every $\phi_n$ belongs to the Sobolev space $H^1(W)$ (this means that both the
kinetic and potential energy of the fluid motion are finite), whereas the set of
these functions restricted to $F$ forms together with a non-zero constant a complete
orthogonal system in $L^2(F)$. It should be emphasised that the most important
eigenfunction is $\phi_1$ because it has the least rate of decay with time caused by
non-ideal effects in real-life fluids.

\section{Vertical circular containers without and with a radial baffle}

In what follows, we use non-dimensional variables chosen so that the container's
radius and the constant acceleration due to gravity are scaled to unity. For this
purpose lengths are scaled to $R$, whereas the velocity potential $\phi$ is scaled
to $(R^3 g)^{1/2}$ (nevertheless, we keep the same notation $\phi$ below); here $R$
and $g$ are the dimensional quantities for the container's radius and the gravity
acceleration respectively. Furthermore, the non-dimensional spectral parameter is
$\nu = R \, \omega^2 / g$ in this case.


\subsection{Vertical circular container without baffle}

In the non-dimensional variables introduced above, the fluid domain under
consideration is $W = \bigl\{(x,y,z): \, x^2 + y^2 < 1, \, z \in (-h,0)\bigr\}$, where $h \in
(0, \infty]$ (note that the case of infinite depth will be also considered). Thus we
have that $F = \bigl\{(x,y,0): \, x^2 + y^2 < 1\bigr\}$ and $\partial W \setminus \overline F =
\overline B \cup S$, where
\[ B = \bigl\{(x,y,-h): \, x^2 + y^2 < 1\bigr\} \quad \mbox{and} \quad S = \bigl\{(x,y,z): \, 
x^2 + y^2 = 1, \, z \in (-h,0)\bigr\} 
\]
are the bottom and the lateral cylindrical surface respectively. Therefore, it is
natural to split the boundary condition (\ref{slosh3}) as follows:
\begin{equation}
\phi_r = 0 \ \ \mbox{on} \ S , \qquad \phi_z = 0 \ \ \mbox{on} \ B
. \label{slosh5-6}
\end{equation}
Here and below $r$ is the first component of the cylindrical coordinates $(r,
\theta, z)$ such that $x = r \cos \theta$ and $y = r \sin \theta$.

The boundary conditions (\ref{slosh5-6}) allow us to separate the
vertical coordinate, thus obtaining the following representations for the velocity
potential and the spectral parameter respectively:
\begin{equation}
\phi (r, \theta, z) = u (r, \theta) \cosh k (z + h) \ \ \mbox{and} \ \ \nu = k \tanh
k h .
\label{u}
\end{equation}
It is clear from the last formula that $\nu$ is an increasing function of $k$.
Moreover, $\nu$ increases with $h$, and so we have
\begin{equation}
\phi (r, \theta, z) = u (r, \theta) \, \E^{k z} \ \ \mbox{and} \ \ \nu = k \ \
\mbox{when} \ h = \infty .
\label{infty}
\end{equation}
Here, as well as in (\ref{u}), $u$ and $k^2$ must be found from the spectral
problem
\begin{equation}
u_{xx} + u_{yy} + k^2 u = 0 \ \ \mbox{in} \ F , \quad u_r (1, \theta) = 0 \ \
\mbox{for} \ \theta \in [0, 2 \pi] , \quad \int_F u r \, \D r \D \theta = 0 .
\label{eigen_0}
\end{equation}

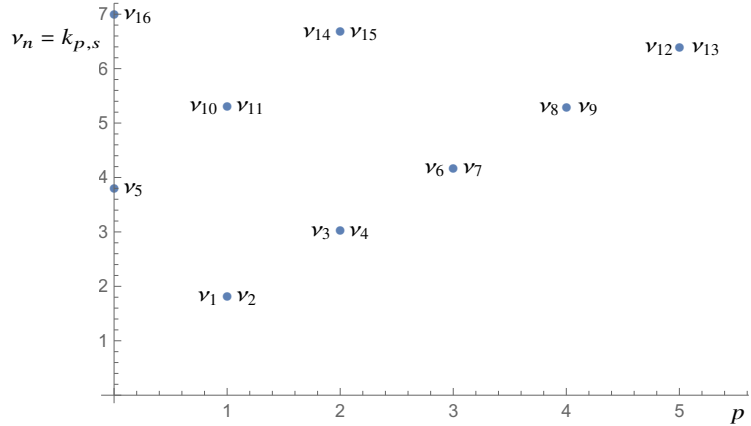
\begin{figure}[t!]
\centering
 \SetLabels
 \L (-0.12*0.9) $\nu_n=k_{p,s}$\\
 \L (0.97*0.01) $p$\\
 \L (0.163*0.29) $\nu_1$~~~$\nu_2$\\
 \L (0.334*0.445) $\nu_3$~~~$\nu_4$\\
 \L (0.05*0.547) $\nu_5$\\
 \L (0.506*0.595) $\nu_6$~~~$\nu_7$\\
 \L (0.68*0.74) $\nu_8$~~~$\nu_9$\\
 \L (0.15*0.742) $\nu_{10}$~~~$\nu_{11}$\\
 \L (0.84*0.882) $\nu_{12}$~~~$\nu_{13}$\\
 \L (0.322*0.918) $\nu_{14}$~~~$\nu_{15}$\\
 \L (0.05*0.96) $\nu_{16}$\\
 \endSetLabels
 \leavevmode\AffixLabels{~\,\includegraphics[height=56mm]{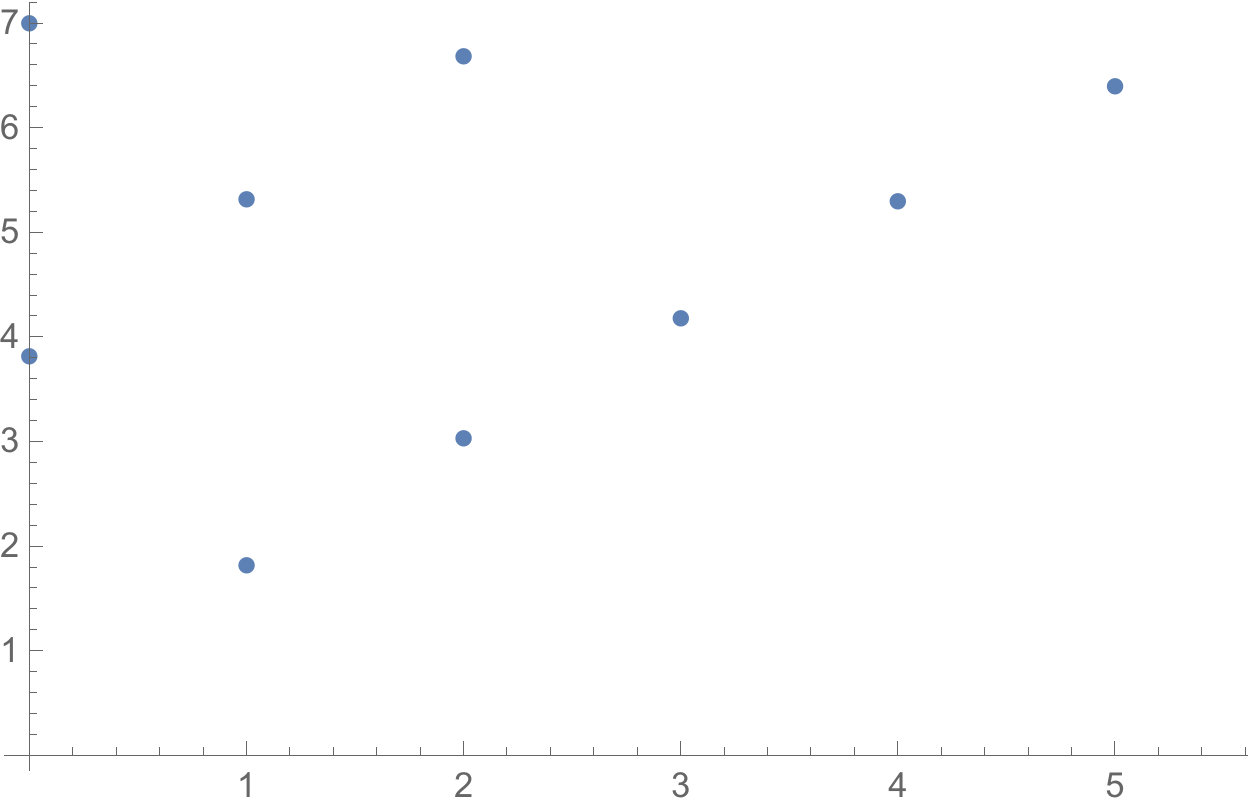}}
 \caption{Values $\nu_n$, $n=1,2,\ldots,16$ in the case of infinite depth.}
 \label{fig:circ_without_baffle}
\end{figure}

It is well known (see, for example, \cite{GN}, \S~3.2) that the set of eigenvalues
of this problem can be written as the following infinite matrix:
\begin{equation}
\bigl( k^2_{m,s} \bigr)_{m=0,s=1}^\infty , \label{matrix}
\end{equation}
where $k_{m,s}$ is the $s$th positive zero of $J'_m$\,---\, the derivative of the
Bessel function $J_m$ (for $m=0$ this numbering differs from that used in \cite{AS},
where $j'_{0,s}$ is the $s$th non-negative zero). Moreover, all eigenvalues are of
multiplicity two when $m \neq 0$; in particular, the lowest eigenvalue is of
multiplicity two as well as the other six corresponding to the eight initial zeroes
of $J'_m$ with various $m$, which are as follows in the increasing order:
\begin{align}
& k_{1,1} = 1.8411... , \ k_{2,1} = 3.0542... , \ k_{0,1} = 3.8317... , \ k_{3,1} =
4.2011... ,  \nonumber\\ & k_{4,1} = 5.3175... , \ k_{1,2} = 5.3314... , \ k_{5,1}
= 6.4156... , \ k_{2,2} = 6.7061... \label{zeroes}
\end{align}
The second formula (\ref{infty}) yields that these values are $\nu_1, \dots ,
\nu_{15}$ for the infinitely deep container and among them only $\nu_5 = k_{0,1}$ is
simple, whereas the next simple eigenvalue is $\nu_{16} = k_{0,2} = 7.0155...$ (the
subsequence of simple eigenvalues is rather sparse); see
Fig.~\ref{fig:circ_without_baffle}.

If $h$ is finite, then we have by the second formula (\ref{u}) that
\begin{equation}
\nu_{m,s} = k_{m,s} \tanh k_{m,s} h , \quad m = 0,1,\dots, \ s = 1,2,\dots \, .
\label{nu}
\end{equation}
In particular, when $h = 2$ (the container's depth is equal to its diameter) the
lowest sloshing eigenvalue is\vspace{-1.5mm}
\begin{equation}
\nu_{1} = k_{1,1} \tanh 2 \, k_{1,1} \approx 1.8387 \, ,
\label{1,1}
\end{equation}
which is sufficiently close to the fundamental eigenvalue shown in
Fig.~\ref{fig:circ_without_baffle}. Thus the fundamental eigenvalue defined by
\eqref{nu} differs from that for infinite depth only for sufficiently small $h$.

The eigenfunctions of problem (\ref{eigen_0}) corresponding to every
$k^2_{m,s}$, $m = 0,1,\dots$, $s = 1,2,\dots$,
are as follows:
\begin{equation}
u_{m,s,1} (r, \theta) = J_m \bigl( k_{m,s} \, r \bigr) \cos m \theta , \quad
u_{m,s,2} (r, \theta) = J_m \bigl( k_{m,s} \, r \bigr) \sin m \theta \, ,
\label{eif}
\end{equation}
where $m \neq 0$ for the functions of the second type. Thus, every eigenfunction can
be written in the form:
\begin{equation}
u_{m,s} (r, \theta) = A J_m \bigl( k_{m,s} \, r \bigr) \cos (m \theta - \beta) \,
. \label{com}
\end{equation}
Here $A$ is an arbitrary non-zero real constant and $\beta$ is a number from $[0, 2
\pi)$ when $m \neq 0$, that is, (\ref{com}) is a non-trivial linear combination of
the two functions (\ref{eif}) in this case; if $m = 0$, then $\cos \beta \neq 0$
can be included into $A$.

From (\ref{u}) and (\ref{infty}), it follows that the elevation of the free surface
is proportional to $\sin (\omega t + \alpha) \, u (x,y)$. Hence the free surface
elevation of every sloshing eigenfunction attains its maxima and minima at the same
points of $\overline F$, where the corresponding function (\ref{com}) does. Also, the
location of nodes on the free surface is defined by the nodal lines of (\ref{com})
in $F$.

Let us consider in detail the properties of
$u_{1,1} (r, \theta) = A J_1 \bigl( k_{1,1} \, r \bigr) \cos (\theta - \beta)$,
corresponding to the lowest eigenvalue $k_{1,1}^2$ and defining the free surface
elevation for the fundamental sloshing mode. To be specific we assume that $A > 0$,
in which case $u_{1,1}$ attains its maximum and minimum at $(1, \beta)$ and $(1, \pi
+ \beta)$ respectively, because $J_1 \bigl( k_{1,1} \, r \bigr)$ monotonically
increases on $(0, 1)$. Since $\beta \in [0, 2 \pi)$ is arbitrary, the maximum can be
attained at any point on $\partial F$, whereas the minimum is attained at the point
symmetric to the point of maximum with respect to the origin. The single nodal line
of $u_{1,1}$ is the diameter orthogonal to that connecting the points of maximum and
minimum. It is worth mentioning that these points for the fundamental sloshing mode
are referred to as `high spots' for the reason discussed in \cite{KK1,KK2}.

Let us turn to properties of the functions $u_{m,s}$ when either $m$ or $s$ is
greater than one. First, for every $u_{m,1}$ all its $m$ maxima and $m$ minima
belong to $\partial F$ and the points, where maxima and minima are attained, are
symmetric with respect to the origin; moreover, each of $m$ nodal lines is a
diameter of $F$. Second, every $u_{1,s}$ has a single maximum and a single minimum
inside the disc bounded by the innermost circular nodal line; the total number of
nodal lines is $s$, one of which is a diameter of $F$, whereas the rest are
circular. Finally, if both $m$ and $s$ are greater than one, then $u_{m,s}$ has
maxima and minima on $\partial F$ as well as inside $F$ and nodal lines of both
types (diameters and circles) exist.

The set of simple eigenvalues of problem (\ref{eigen_0}) is $\bigl\{ k_{0,s}^2
\bigr\}_{s=1}^\infty$, and the axisymmetrc eigenfunction $u_{0,s} (r, \theta) = A J_m
\bigl( k_{0,s} \, r \bigr)$ with $A \neq 0$ corresponds to each of these
eigenvalues. If $A > 0$, then the global maximum is attained at the origin and the
global minimum is attained at every point of the circumference $r = k_{0,1} /
k_{0,s}$, whereas the points of $\partial F$ deliver a negative (positive) local
minimum (maximum respectively) when $s > 1$ is odd (even respectively). The
eigenfunction $u_{0,s}$ has $s$ nodal lines whose equations are $r = j_{0,n} /
k_{0,s}$, $n = 1,\dots,s$, and $j_{0,n}$ is the $n$th positive zero of $J_0$.

In particular, the eigenfunction $u_{0,1} (r, \theta) = A J_m \bigl( k_{0,1} \, r
\bigr)$ (it corresponds to the smallest simple eigenvalue for which $k_{0,1} =
3.8317...$) monotonically decreases with $r$ and the equation of its single nodal
line is $r = j_{0,1} / k_{0,1} \approx 0.6276$, where $j_{0,1} = 2.4048...$ is the
least positive zero of $J_0$.

\subsection{Vertical circular container with a radial baffle}

We assume that the second container is the same as above, but complemented by the
rectangular rigid baffle $L = \{(r, 0, z): \, r \in [0, 1] , \, z \in [-h,0]\}$,
that is, the fluid domain is as follows:
\[ \bar W = \{(r, \theta, z): \, r \in (0, 1) , \, \theta \in (0, 2 \pi) , \, z \in
(-h,0)\} .
\]
Thus, $\partial \bar W$\,---\,the boundary of the fluid domain\,---\,is the following
union of surfaces $\skew4\overline{\bar F} \cup L \cup \skew4\overline{\bar B} \cup \bar S$.
Here, the free surface $\bar F$ and the container's bottom $\bar B$ are $F$ and
$B$, respectively, from \S\,2.1 cut by the top and bottom side of $L$ respectively,
whereas the lateral cylindrical surface $\bar S$ is $S$ from \S\,2.1 cut vertically
along the interval $\{ r = 1 , \, \theta = 0 , \, z \in (-h,0)\}$. Here and below the accent $\bar{~}$ is
used to distinguish notations from those for the container without baffle.

On both sides of $L$, the no-flow condition must hold, that is, the equalities
\begin{equation}
\phi_\theta (r, 0, z) = \phi_\theta (r, 2 \pi, z) = 0 \ \ \mbox{for} \ r \in (0, 1)
, \ z \in (-h,0)
\label{slosh7}
\end{equation}
complement the boundary conditions (\ref{slosh2}) and (\ref{slosh5-6}).

Representations (\ref{u}) and (\ref{infty}) are still valid because they depend only
on the fact that the bottom is horizontal and on its depth. However, taking into
account the boundary conditions (\ref{slosh7}), instead of (\ref{eigen_0}), we obtain the spectral problem
\begin{align}
& u_{xx} + u_{yy} + k^2  = 0 \ \ \mbox{in} \ \bar F, \quad u_r (1, \theta) = 0 \
\ \mbox{for} \ \theta \in (0, 2 \pi) , \label{eigen_1} \\ & u_\theta (r, 0) =
u_\theta (r, 2 \pi) = 0 \ \ \mbox{for} \ r \in (0, 1) , \quad \int_{\bar F} u  r\,
\D r \D \theta = 0. 
\label{eigen_2}
\end{align} 

All eigenvalues of this problem are simple and they can be written in the matrix
form similar to (\ref{matrix}), but now the elements are equal to $\bar
k_{m/2,s}^2$, $m = 0,1,\dots , \ s = 1,2,\dots$ (see \cite{GN}, \S~3.2). Here $\bar
k_{p,s}$ is the $s$th positive zero of $J'_p$\,---\,the derivative of the Bessel
function $J_p$ with integer or half-integer $p \geq 0$ (again our notation differs
from that in \cite{AS} for $m=0$). The twenty initial values in the increasing order
are as follows:
\begin{align}
& \bar k_{1/2,1} = 1.1655... , \ \bar k_{1,1} = 1.8411... , \ \bar k_{3/2,1} =
2.4605... , \ \bar k_{2,1} = 3.0542... , \nonumber \\ & \bar k_{5/2,1} = 3.6327...
, \ \bar k_{0,1} = 3.8317... , \ \bar k_{3,1} = 4.2011... , \ \bar k_{1/2,2} =
4.6042... , \nonumber \\ & \bar k_{7/2,1} = 4.7621... , \ \bar k_{4,1} = 5.3175...
, \ \bar k_{1,2} = 5.3314... , \ \bar k_{9/2,1} = 5.8684... , \nonumber \\ & \bar
k_{3/2,2} = 6.0292... , \ \bar k_{5,1} = 6.4156... , \ \bar k_{2,2} = 6.7061... , \
\bar k_{11/2,1} = 6.9597... , \nonumber \\ & \bar k_{0,2} = 7.0155... , \ \bar
k_{5/2,2} = 7.3670... , \ \bar k_{6,1} = 7.5012... , \ \bar k_{1/2,3} = 7.7898...
\label{zeroes_b}
\end{align}
%
%
It should be noted that $k_{1,1}$\,---\,the first value in
(\ref{zeroes})\,---\,coincides with $\bar k_{1,1}$ which is second here, whereas the
eighth value in (\ref{zeroes}) is only fifteenth here. It is clear that every
eigenvalue $\bar k_{m/2,s}^2$ with even $m$ is also an eigenvalue of problem
(\ref{eigen_0}), but for eigenvalues with odd $m$ this is not true.

\begin{figure}[t!]
\centering
 \SetLabels
 \L (-0.12*0.9) $\bar\nu_n=\bar k_{p,s}$\\
 \L (1.02*0.02) $p$\\
 \L (0.123*0.185) $\bar\nu_1$\\
 \L (0.204*0.262) $\bar\nu_2$\\
 \L (0.283*0.33) $\bar\nu_3$\\
 \L (0.365*0.3975) $\bar\nu_4$\\
 \L (0.443*0.464) $\bar\nu_5$\\
 \L (0.05*0.487) $\bar\nu_6$\\
 \L (0.523*0.528) $\bar\nu_7$\\
 \L (0.124*0.575) $\bar\nu_8$\\
 \L (0.598*0.592) $\bar\nu_9$\\
 \L (0.679*0.655) $\bar\nu_{10}$\\
 \L (0.206*0.655) $\bar\nu_{11}$\\
 \L (0.755*0.715) $\bar\nu_{12}$\\
 \L (0.282*0.734) $\bar\nu_{13}$\\
 \L (0.837*0.778) $\bar\nu_{14}$\\
 \L (0.363*0.815) $\bar\nu_{15}$\\
 \L (0.914*0.839) $\bar\nu_{16}$\\
 \L (0.047*0.845) $\bar\nu_{17}$\\
 \L (0.438*0.887) $\bar\nu_{18}$\\
 \L (0.995*0.902) $\bar\nu_{19}$\\
 \L (0.125*0.934) $\bar\nu_{20}$\\
 \endSetLabels
 \leavevmode\AffixLabels{ \includegraphics[height=56mm]{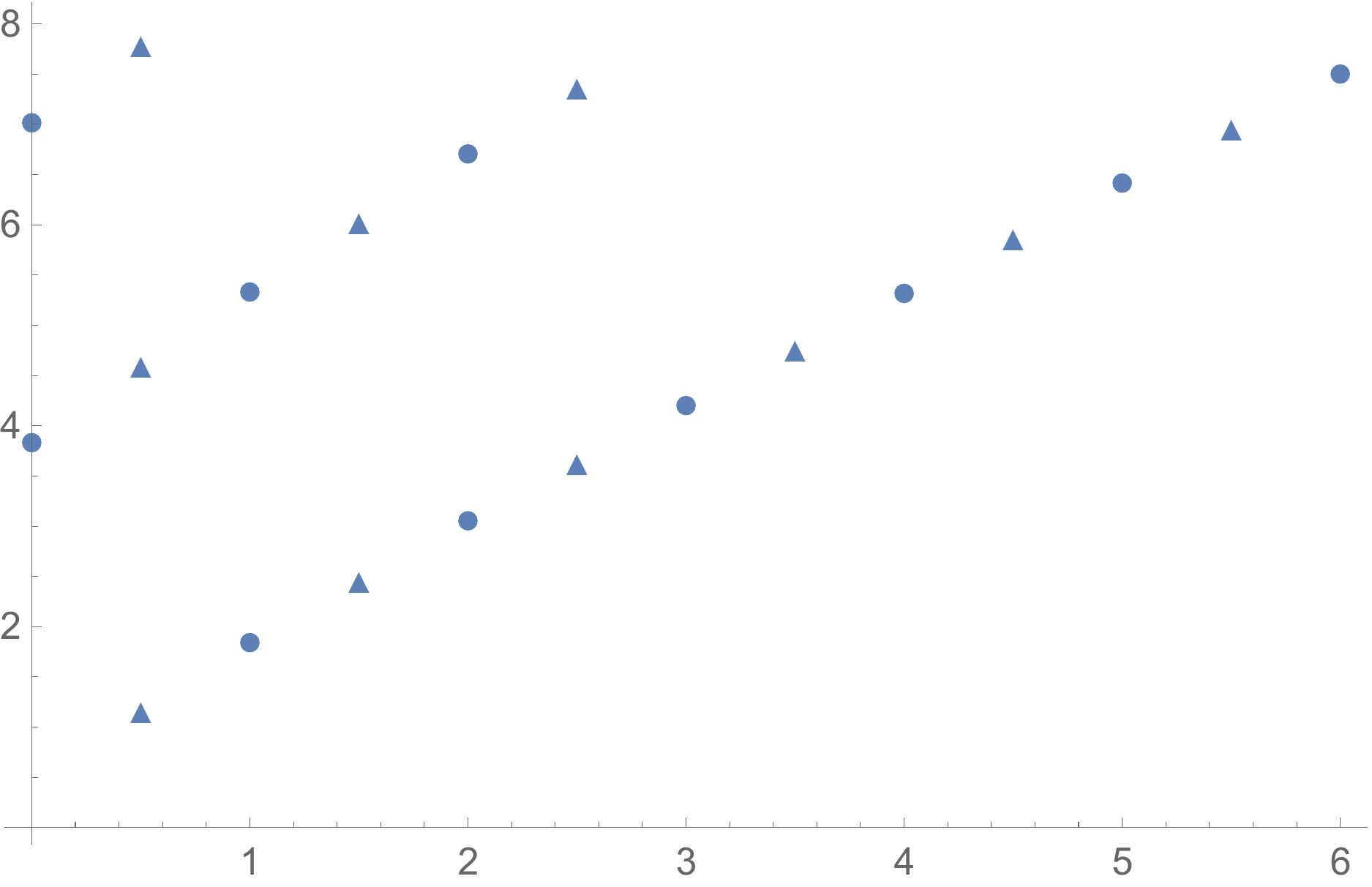}}
 \caption{Values $\bar\nu_n$, $n=1,2,\ldots,20$, for a circular container with baffle in the case of infinite depth.}
 \label{fig:circ_with_baffle}
\end{figure}
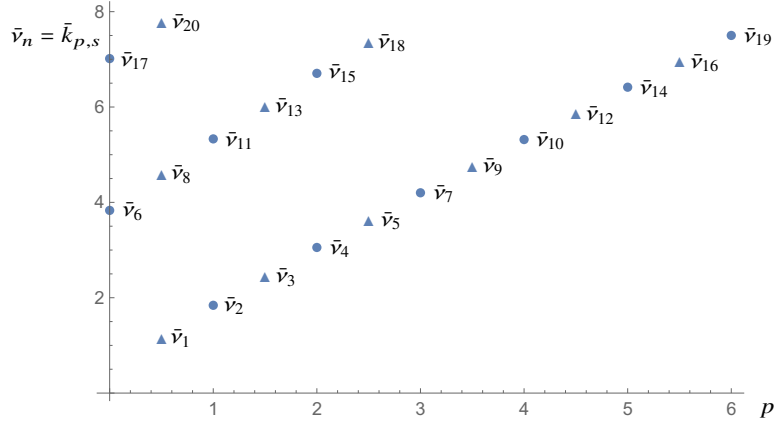

As in \S~2.1, the second formula (\ref{infty}) yields that the numbers
(\ref{zeroes_b}) are the eigenvalues $\bar \nu_1, \dots , \bar\nu_{20}$ (shown in Fig.~\ref{fig:circ_with_baffle}) for the
infinitely deep container with a radial baffle.
In this case, the ratios $\nu_1 / \bar\nu_1 , \dots , \nu_{16} / \bar\nu_{16}$
are as follows:
\begin{align*}
&  1.5796... , \ 1 , \ 1.2412... , \ 1 , \ 1.0547... , \ 1.0964... , \ 1 , \
1.1549... , \\ & 1.1166... , \ 1.0026... , \ 1 , \ 1.0932... , \ 1.0640... , \
1.0452... , \ 1 , \ 1.0080... 
\end{align*}
Since most of the eigenvalues have multiplicity two in the absence of baffle, we see
that $\nu_n = \bar\nu_n$ for some $n$, but, in general, the behaviour of $\nu_n /
\bar\nu_n$ demonstrates no regular pattern, at least for this initial part of the
sequence.

\begin{figure}[t!]
\centering
 \SetLabels
 \L (-0.09*0.96) $\bar k_{p,s},$\\
 \L (-0.09*0.88) $\bar\nu_{p,s}$\\
 \L (0.97*0.01) $p$\\
 \endSetLabels
 \leavevmode\AffixLabels{ \includegraphics[height=56mm]{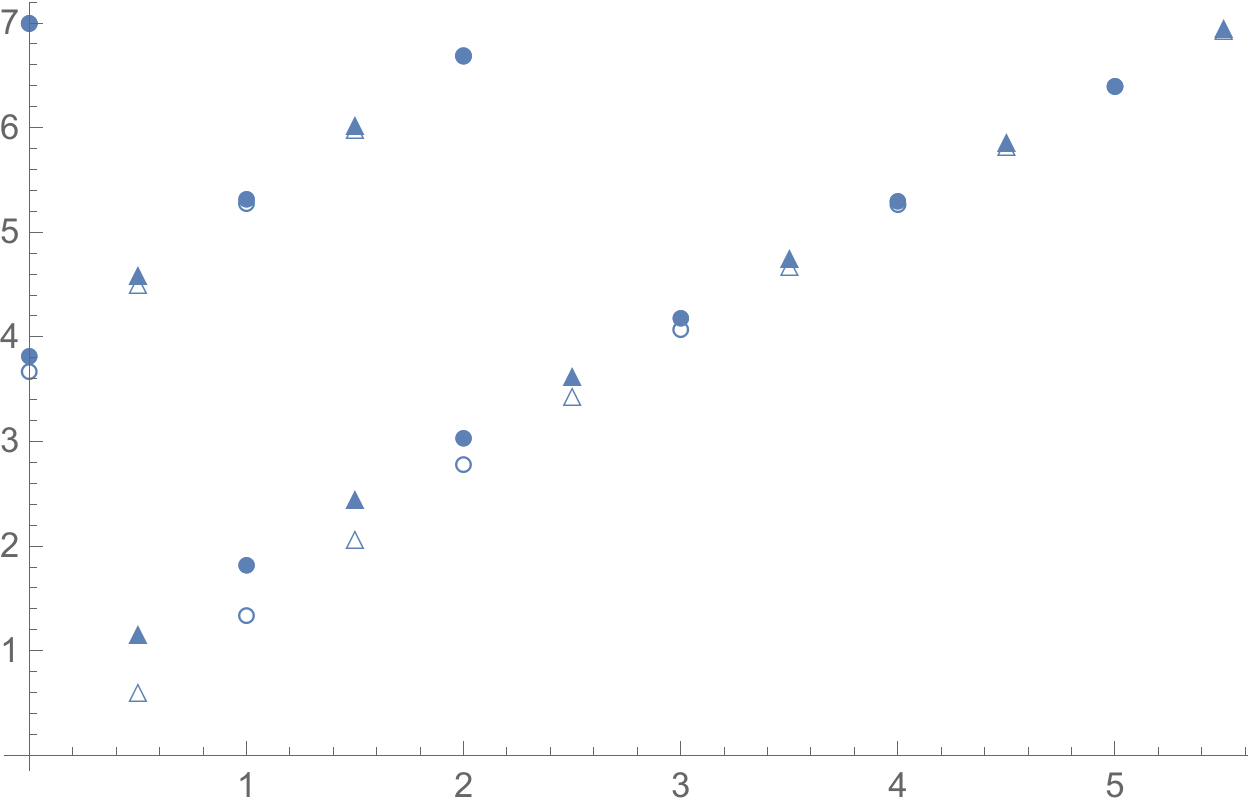}}
 \caption{Comparison of $k_{p,s}$ (disks), $\bar k_{p,s}$ (disks and filled triangles)\,---\,corresponding to the infinite depth case\,---\,and $k_{p,s}\tanh hk_{p,s}$ (circles), $\bar k_{p,s}\tanh h\bar k_{p,s}$ (circles and triangles) for the finite depth $h = 0.5$.}
 \label{fig:comp}
\end{figure}

If $h$ is finite, then the second formula (\ref{u}) gives that
\begin{equation*}
\bar\nu_{m/2,s} = \bar k_{m/2,s} \tanh \bar k_{m/2,s} h , \quad m = 0,1,\dots \, ,
\ s = 1,2,\dots
\label{nu_b}
\end{equation*}
To show how this formula distinguishes from (\ref{nu}) we take \mbox{$h = 2$}\,---\,the
same container's depth as in (\ref{1,1})\,---\,which gives for $m=1$:
\begin{equation*}
\bar\nu_{1} = \bar k_{1/2,1} \tanh 2 \, \bar k_{1/2,1} \approx 1.1436 \, .
\label{1/2,1}
\end{equation*}
Thus, $\bar\nu_{1} \approx \nu_{1} \cdot 0.6220$, and so the presence of the
radial baffle in this circular container substantially diminishes the lowest
sloshing eigenvalue comparing with the same container without baffle.
Comparison of $\bar k_{p,s}$ and $\bar\nu_{p,s}$ for $h = 0.5$ is done in Fig.~\ref{fig:comp}.

In view of conditions imposed on $u_\theta$ on both sides of the baffle, problem
(\ref{eigen_1}), (\ref{eigen_2}) has only one eigensolution corresponding to the
eigenvalue $\bar k_{m/2,s}^2$, namely
\begin{equation}
\bar u_{m,s} (r, \theta) = J_{m/2} \left( \bar k_{m/2,s} \, r \right) \cos \frac{m
\theta}{2} , \quad m=0,1,2,\dots,\ s=1,2,\dots 
\label{eifb}
\end{equation}
Its properties concerning maxima, minima and nodal lines are absolutely different
from those of $u_{m,s}$; see formula (\ref{com}). Let us consider these properties
for the fundamental eigenfunction
\[ \bar u_{1,1} (r, \theta) = J_{1/2} \left( \bar k_{1/2,s} \, r \right) \cos 
\frac{\theta}{2} \, .
\]
It is an odd function of $y$, and so the unit interval of the negative $x$-axis is
its nodal line. Furthermore, $\bar u_{1,1} (1, 0)$ and  $\bar u_{1,1} (1, 2 \pi)$
are the only maximum and minimum values of this function. The fact that the `high
spots' are the points $(1, 0)$ and $(1, 2 \pi)$ adjacent to the baffle $L$ is of
practical importance because it is easier to suppress extremal sloshing localised
at a particular place.

Turning to the case when either $m$ or $s$ is greater than one, we first consider
$\bar u_{m,1}$ with $m \geq 2$. All $m$ maxima and $m$ minima of this function
belong to
\[ \{ (r, \theta): \, r = 1 , \, \theta \in [0, 2 \pi] \} \, .
\]
Moreover, if $m$ is odd, then $\bar u_{m,1}$ is an odd function of $y$, and $(1,
0)$ and $(1, 2 \pi)$ are the points, where maximum and minimum, respectively, are
attained by this function. Its other points of maxima and minima are also symmetric
about the $x$-axis. All nodal lines of $\bar u_{m,1}$ with odd $m$ are diameters of
the unit disc with exception for the unit interval of the negative $x$-axis.

In the case of even $m$, formula (\ref{eifb}) yields that $\bar u_{m,1}$ is an even
function of $y$, and both $(1, 0)$ and $(1, 2 \pi)$ are points of maximum for this
function. Other points of maximum are also symmetric about the $x$-axis and the same
is true for the points of minimum. All nodal lines of $\bar u_{m,1}$ with even $m$
are diameters of the unit disc.

Properties of $\bar u_{1,s}$ with $s \geq 2$ and $\bar u_{m,s}$ when both $m$ and
$s$ are greater than one are similar to those of $u_{1,s}$ with $s \geq 2$ and
$u_{m,s}$ respectively.

As in \S~2.1, every sloshing eigenfunction has maxima and minima of its free surface
elevation at the same points of $\skew4\overline{\bar F}$, where the corresponding
function (\ref{eifb}) does. Also, the location of nodes on the free surface is
defined by the nodal lines of functions (\ref{eifb}) in $\bar F$.


\section{Vertical annular containers without and with a radial baffle}

In this section, we use non-dimensional variables chosen so that the container's
exterior radius and the constant acceleration due to gravity are scaled to unity
(see \S\,2 for details).


\subsection{Vertical annular container without baffle}

Let $\rho$ denote the non-dimensional radius of the inner wall coaxial with the
exterior one, then the fluid domain under consideration is 
\[ W^{\ann} = \bigl\{(x,y,z): \, \rho < x^2 + y^2 < 1, \, z \in (-h,0)\bigr\} , \quad \mbox{where}
\ h \in (0, \infty] .
\]

Representations (\ref{u}) and (\ref{infty}) are valid because they depend only
on the fact that the bottom is horizontal and on its depth. Taking into
account the boundary conditions (\ref{slosh7}), we obtain the spectral problem
\begin{align}
& u_{xx} + u_{yy} + k^2  = 0 \ \ \mbox{in} \ F^{\ann}=\bigl\{(x,y): \, \rho < x^2 + y^2 < 1\bigr\}, \label{eigen_1-a}\\
&u_r (1, \theta) = u_r (\rho, \theta) = 0 \
\ \mbox{for} \ \theta \in (0, 2 \pi), \quad \int_{F^{\ann}} u r
\,\D r \D \theta = 0.
\label{eigen_2-a}
\end{align}

Using \cite{GN} we write the solutions to \eqref{eigen_1-a}, \eqref{eigen_2-a} as follows:
\begin{equation}
u^{\ann}_{m,s}(r,\theta)=\frac{J_m(k^{\ann}_{m,s}r)Y'_m(k^{\ann}_{m,s})-J'_m(k^{\ann}_{m,s})Y_m(k^{\ann}_{m,s}r)}{J_m(k^{\ann}_{m,s})Y'_m(k^{\ann}_{m,s})-J'_m(k^{\ann}_{m,s})Y_m(k^{\ann}_{m,s})}
\begin{cases}
\cos m\theta,\\
\sin m\theta\ \ (m\neq0),
\end{cases}
\label{eq:uwo}
\end{equation}
where $m=0,1,2,\ldots$, $J_m(\cdot)$ and $Y_m(\cdot)$ are Bessel's functions of the first and the second kind. The denominator in \eqref{eq:uwo} is introduced to normalize the radial factor to unity at $r=1$. The values $k^{\ann}_{m,s}$, for a fixed $m$ and $s=1,2,3,\ldots$, are the increasing roots of the equation
\begin{equation}
J'_m(k^{\ann}_{m,s})Y'_m(\rho k^{\ann}_{m,s})-J'_m(\rho k^{\ann}_{m,s})Y'_m(k^{\ann}_{m,s})=0.
\label{eq:crossproduct1}
\end{equation}
For the properties of the cross-product in the left-hand side, see, e.g., \cite{Gotlieb85,Sorolla2013} and references therein.

\begin{figure}[t!]
\centering
 \SetLabels
 \L (-0.06*0.96) $k^{\ann}_{m,s}$,\\
 \L (0.99*0.01) $m$\\
 \endSetLabels 
 \leavevmode\AffixLabels{ \includegraphics[height=56mm]{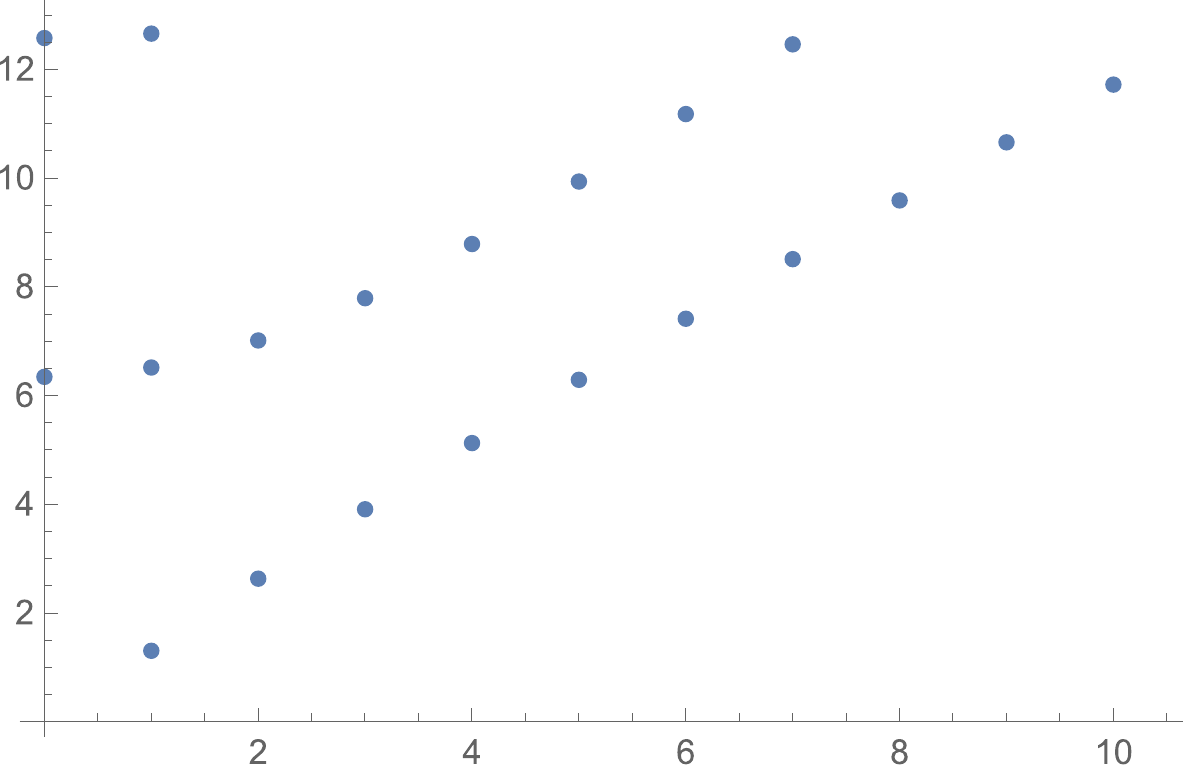}}
 \caption{Values $k^{\protect\ann}_{m,s}$ for an annular container of infinite depth without baffle,  $\rho=1/2$.}
 \label{fig:ann0}
\end{figure}

A set of values $k^{\ann}_{m,s}$, computed for $\rho=1/2$, is shown in Fig.~\ref{fig:ann0} and is given here in the ascending order:
\begin{align*}
&k^{\ann}_{1,1}=1.3546...,\ 
k^{\ann}_{2,1}=2.6812...,\
k^{\ann}_{3,1}=3.9577...,\
k^{\ann}_{4,1}=5.1752...,\\&
k^{\ann}_{5,1}=6.3388...,\
k^{\ann}_{0,1}=6.3931...,\
k^{\ann}_{1,2}=6.5649...,\
k^{\ann}_{2,2}=7.0625...,\\&
k^{\ann}_{6,1}=7.4621...,\
k^{\ann}_{3,2}=7.8401...,\
k^{\ann}_{7,1}=8.5586...,\
k^{\ann}_{4,2}=8.8364...,\\&
k^{\ann}_{8,1}=9.6382...,\
k^{\ann}_{5,2}=9.9858...,\
k^{\ann}_{9,1}=10.7070...,\
k^{\ann}_{6,2}=11.2269...,\\&
k^{\ann}_{10,1}=11.7688...,\
k^{\ann}_{7,2}=12.5094...,\
k^{\ann}_{0,2}=12.6246...,\
k^{\ann}_{1,3}=12.7064...
\end{align*}


It should be noted at the point that all eigenvalues $\bigl[k^{\ann}_{m,s}\bigr]^2$ for $m,s=1,2,\ldots$ have multiplicity two, whereas $\bigl[k^{\ann}_{0,s}\bigr]^2$ for $s=1,2,\ldots$ are simple.

\vspace{-4mm}

\subsection{Vertical annular container with a radial baffle}

\vspace{-2mm}

In the section we assume that the container is the same as above, but complemented by the
rectangular rigid baffle $L = \{(r, 0, z): \, r \in [\rho, 1] , \, z \in [-h,0]\}$,
that is, the fluid domain is as follows:
\[ \bar{W}^{\annB} = \{(r, \theta, z): \, r \in (\rho, 1) , \, \theta \in (0, 2 \pi) , \, z \in
(-h,0)\} .
\]


On both sides of $L$, the no-flow condition must hold, that is, the equalities
\begin{equation*}
\phi_\theta (r, 0, z) = \phi_\theta (r, 2 \pi, z) = 0 \ \ \mbox{for} \ r \in (\rho, 1)
, \ z \in (-h,0)
\label{slosh7-a}
\end{equation*}
complement the boundary conditions.

Representations (\ref{u}) and (\ref{infty}) are still valid. Taking into
account the boundary conditions (\ref{slosh7}), we obtain the spectral problem
\begin{align}
& u_{xx} + u_{yy} + k^2  = 0 \ \ \mbox{in} \ \bar{F}^{\annB}, \quad u_r (1, \theta) = u_r (\rho, \theta) = 0 \
\ \mbox{for} \ \theta \in (0, 2 \pi) , \label{eigen_1-b} \\ & u_\theta (r, 0) =
u_\theta (r, 2 \pi) = 0 \ \ \mbox{for} \ r \in (\rho, 1) , \quad \int_{\bar{F}^{\annB}} u r
\,\D r \D \theta = 0 .
\label{eigen_2-b}
\end{align}


\begin{figure}[t!]
\centering
 \SetLabels
 \L (-0.08*0.96) $k^{\ann}_{p,s}$,\\
 \L (-0.08*0.88) $\bar{k}^{\annB}_{p,s}$\\
 \L (1*0.01) $p$\\
 \endSetLabels 
 \leavevmode\AffixLabels{ \includegraphics[height=56mm]{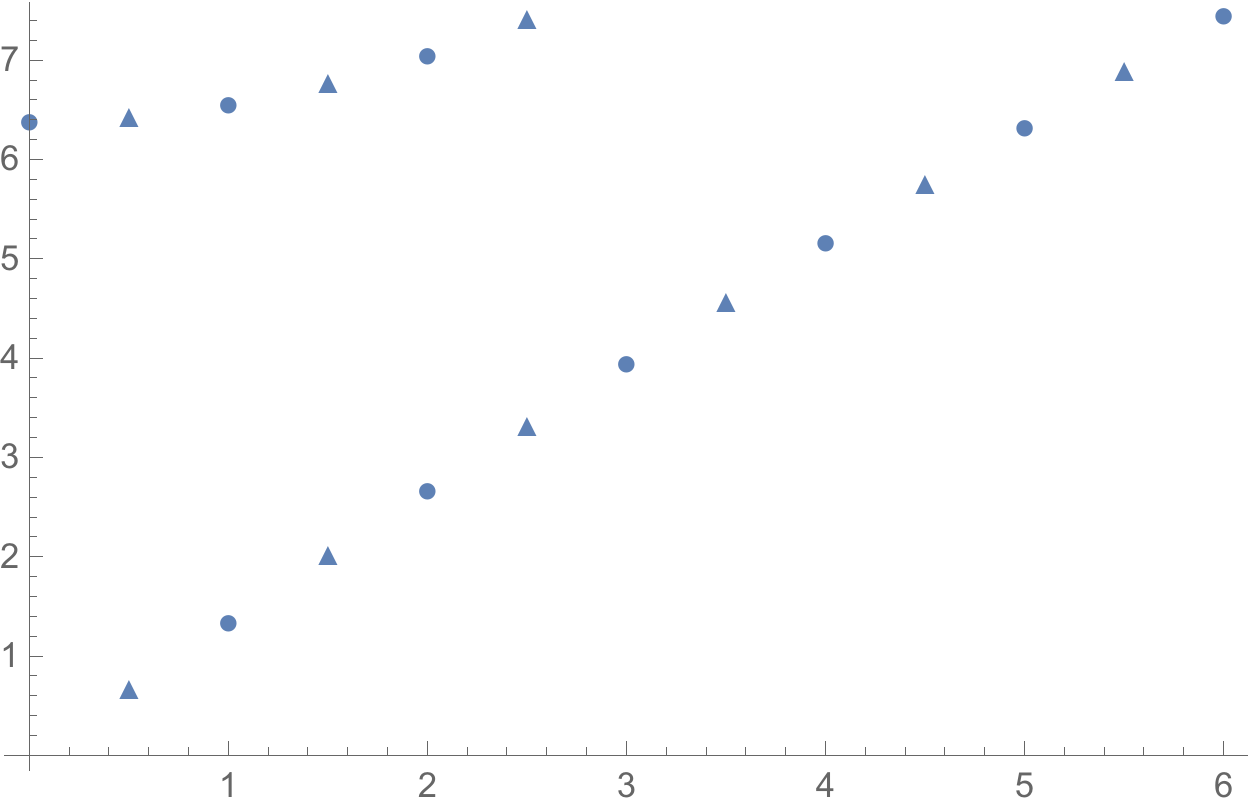}}
 \caption{Values $k^{\protect\ann}_{p,s}$ (disks) for an annular container without baffle, $\rho=1/2$, and 
 values $\bar{k}^{\annB}_{p,s}$ (disks and filled triangles) for an annular container with baffle, $\rho=1/2$.}
 \label{fig:ann}
\end{figure}

All eigenvalues of this problem, namely, $\bigl[\bar{k}^{\annB}_{m/2,s}\bigr]^2$, $m = 0,1,2,\ldots $, $s = 1,2,\ldots$,  are simple;
we can write the solutions to \eqref{eigen_1-b}, \eqref{eigen_2-b} as follows:
\[
\bar{u}^{\annB}_{m,s}(r,\theta)=R_{m/2,s}(r)
\cos \frac{m}{2}\theta,
\]
where $R_{m/2,s}(r)=\mathring{R}_{m/2,s}(r)/\mathring{R}_{m/2,s}(1)$  (normalized to 1 at $r=1$),
\[
\mathring{R}_{m/2,s}(r)=J_{m/2}\bigl(\bar{k}^{\annB}_{m/2,s}r\bigr)Y'_{m/2}\bigl(\bar{k}^{\annB}_{m/2,s}\bigr)-
J'_{m/2}\bigl(\bar{k}^{\annB}_{m/2,s}\bigr)
Y_{m/2}\bigl(\bar{k}^{\annB}_{m/2,s}r\bigr),
\]
$J_{m/2}(\cdot)$ and $Y_{m/2}(\cdot)$ are Bessel's functions of the first and the second kind. Here the values $\bar{k}^{\annB}_{m/2,s}$, for a fixed $m$ and $s=1,2,3,\ldots$, are the increasing roots of the equation
\begin{equation}
J'_{m/2}\bigl(\bar{k}^{\annB}_{m/2,s}\bigr)Y'_{m/2}\bigl(\rho \bar{k}^{\annB}_{m/2,s}\bigr)-J'_{m/2}\bigl(\rho \bar{k}^{\annB}_{m/2,s}\bigr)Y'_{m/2}\bigl(\bar{k}^{\annB}_{m/2,s}\bigr)=0.
\label{eq:crossproduct2}
\end{equation}

\begin{figure}[t!]
\centering
 \SetLabels
 \L (-0.09*0.9) $\frac{\bar{k}^{\annB}_{1/2,1}}{k^{\ann}_{1,1}}$\\
 \L (0.89*-0.01) $\rho$\\
 \endSetLabels
 \leavevmode\AffixLabels{ \includegraphics[height=50mm]{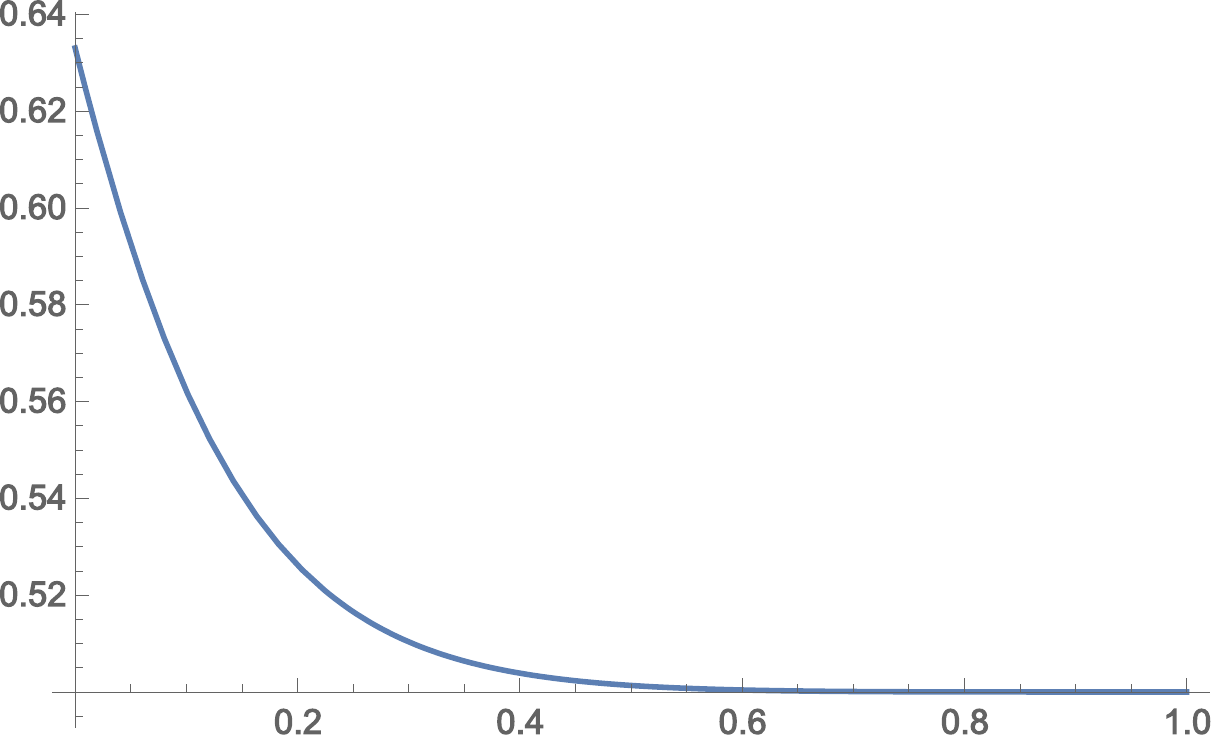}}
 \caption{Ratio $\bar{k}^{\annB}_{1/2,1}/k^{\ann}_{1,1}$ (corresponding to the lowest sloshing eigenvalues) for annular container with and without baffle. Dependence on the internal radius $\rho$.}
 \label{fig:ratio}
\end{figure}
%
Some initial values for $\rho=1/2$ (see Fig.~\ref{fig:ann}) in the increasing order are as follows:
\begin{align*}
&
\bar{k}^{\annB}_{1/2,1}=0.6791...,\
\bar{k}^{\annB}_{1,1}=1.3546...,\
\bar{k}^{\annB}_{3/2,1}=2.0230...,\
\bar{k}^{\annB}_{2,1}=2.6812...,\\&
\bar{k}^{\annB}_{5/2,1}=3.3266...,\
\bar{k}^{\annB}_{3,1}=3.9577...,\
\bar{k}^{\annB}_{7/2,1}=4.5738...,\
\bar{k}^{\annB}_{4,1}=5.1752...,\\&
\bar{k}^{\annB}_{9/2,1}=5.7630...,\
\bar{k}^{\annB}_{5,1}=6.3388...,\
\bar{k}^{\annB}_{0,1}=6.3931...,\
\bar{k}^{\annB}_{1/2,2}=6.4363...,\\&
\bar{k}^{\annB}_{1,2}=6.5649...,\
\bar{k}^{\annB}_{3/2,2}=6.7754...,\
\bar{k}^{\annB}_{11/2,1}=6.9046...,\
\bar{k}^{\annB}_{2,2}=7.0625...,\\&
\bar{k}^{\annB}_{5/2,2}=7.4199...,\
\bar{k}^{\annB}_{6,1}=7.4621...
\end{align*}


\begin{figure}[t!]
\centering
 \SetLabels
 \L (-0.07*0.96) $k^{\ann}_{p,s}$,\\
 \L (-0.07*0.88) $\bar{k}^{\annB}_{p,s}$\\
 \L (0.99*0.01) $p$\\
 \endSetLabels
 \leavevmode\AffixLabels{ \includegraphics[height=56mm]{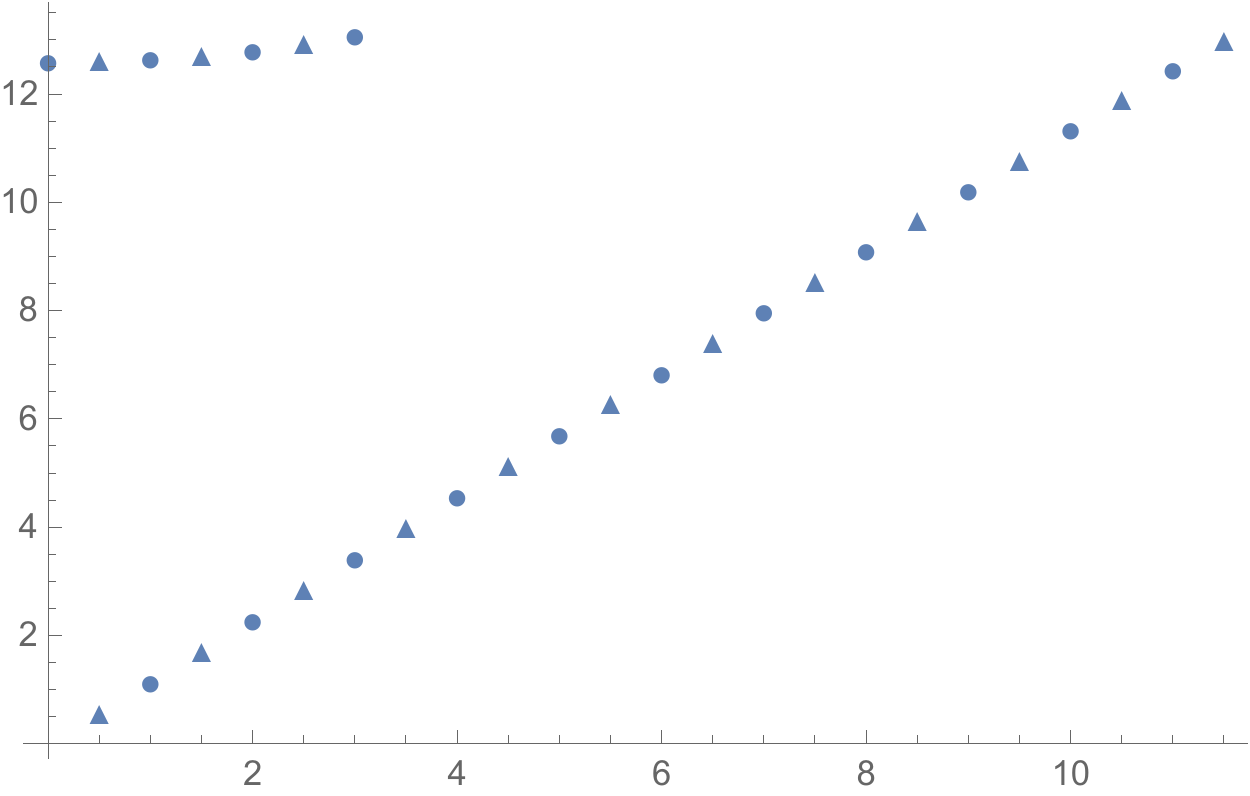}}
 \caption{Values $k^{\protect\ann}_{p,s}$ (disks) for annular container without baffle, $\rho=3/4$, and 
 values $\bar{k}^{\annB}_{p,s}$ (disks and filled triangles) for  annular container with baffle, $\rho=3/4$.}
 \label{fig:ann2}
\end{figure}

It is important to note that the presence of the baffle in the annular container also substantially diminishes the lowest sloshing eigenvalue comparing with the same container without baffle. Figure~\ref{fig:ratio} shows the dependence of the ratio $\bar{k}^{\annB}_{1/2,1}/k^{\ann}_{1,1}$ on the internal radius $\rho$. 
It can also be noted that the fundamental mode is expressed in a fairly simple form:
\[
\bar{u}^{\annB}_{1,1}(r,\theta)=\frac{1}{\sqrt{r}}\Bigl[
\cos\bigl(\bar{k}^{\annB}_{1/2,1}(1-r)\bigr)-(2\bar{k}^{\annB}_{1/2,1})^{-1}\sin\bigl(\bar{k}^{\annB}_{1/2,1}(1-r)\bigr)
\Bigr]
\cos \frac{\theta}{2}.
\]


Changes of $k^{\ann}_{m,s}$ and $\bar{k}^{\annB}_{m,s}$ as the annular container becomes thinner can be observed in Fig.~\ref{fig:ann2}, computed for $\rho=3/4$. Using results of \cite{McMahon1894} and \cite{Buchholz49} on the asymptotic behaviour of the cross-product appearing in the left-hand side of \eqref{eq:crossproduct1} and \eqref{eq:crossproduct2}, it can be shown that  $\bar{k}^{\annB}_{p,1}\to p$ as $\rho\to1^-$, whereas $\bar{k}^{\annB}_{p,s}\to \infty$ for $s=2,3,\ldots$


\begin{figure}[t!]
\centering
 \SetLabels
 \L (-0.08*0.9) $R_{p,s}(r)$\\
 \L (1.02*0.515) $r$\\
  \L (0.39*0.75) $(i)$\\
  \L (0.18*0.19) $(ii)$\\
  \L (0.29*0.11) $(iii)$\\
  \L (0.08*0.86) $(iv)$\\
 \endSetLabels
 \leavevmode\AffixLabels{ \includegraphics[height=52mm]{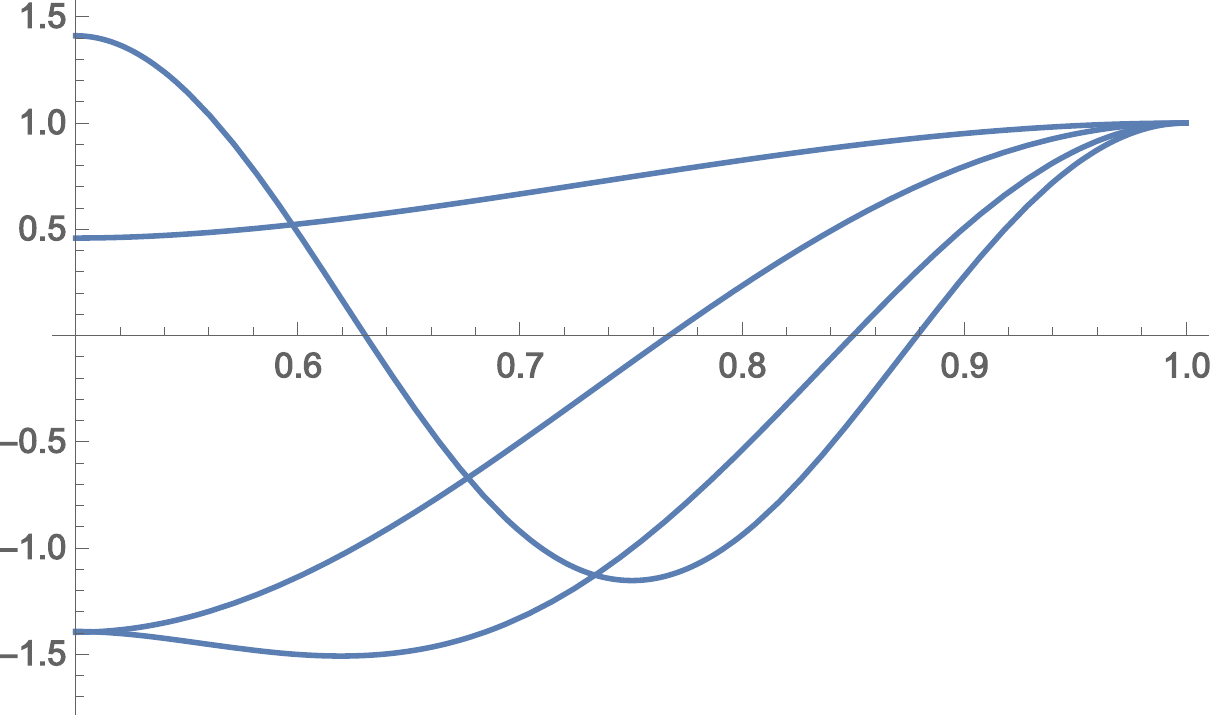}}
 \caption{Functions $R_{p,s}(r)$ for $\rho=1/2$ and $\{p,s\}=\{4, 1\}$ (i), $\{p,s\}=\{0, 1\}$ (ii), $\{p,s\}=\{7, 2\}$ (iii), $\{p,s\}=\{1/2, 3\}$ (iv).}
 \label{fig:spec_rad}
\end{figure}
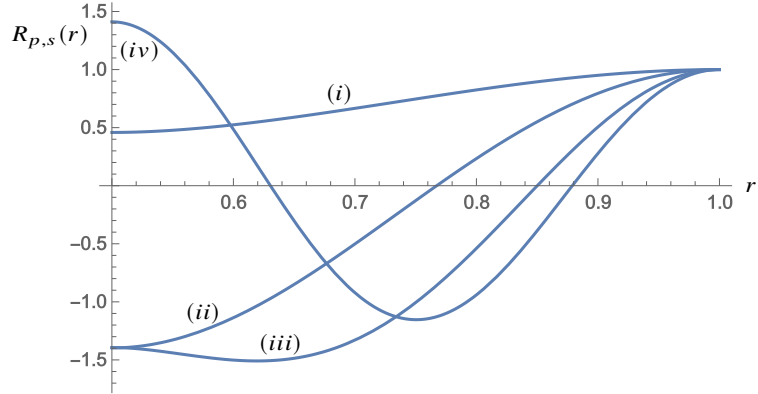

Of interest is the position of extrema of the radial component of the eigenfunction $R_{m/2,s}(r)$
as $r\in[\rho,1]$. In Fig.~\ref{fig:spec_rad} some specific forms of $R_{m/2,s}(r)$ are presented for $\rho=1/2$ to demonstrate different positions of the `high spots'. The curve (i) corresponds to $\bar{k}^{\annB}_{4, 1}=5.1752...$ and the maximum is located at the point, where the baffle
is attached to the outer side wall ($r=1$). The curve (ii) corresponds to $\bar{k}^{\annB}_{0, 1}=6.3931...$ and the minimum is located at the point, where the baffle
is attached to the inner side wall ($r=\rho$). The curve (iii) corresponds to $\bar{k}^{\annB}_{7, 2}=12.5094...$ and shows the case when the maximum of absolute value is located at an inner point of the baffle. The curve (iv) corresponds to $\bar{k}^{\annB}_{1/2, 3}=12.6451...$ and shows more complicated behaviour of the computed eigenfunctions as $s$ increases.


\section{Discussion}

In $\S$~3, we analysed solutions of two spectral problems one of which describes sloshing in a
vertical circular container of constant (possibly infinite) depth, whereas the container considered in
the other problem apart from the same bottom and side wall has also the vertical
baffle that goes from the free surface to the bottom and connects the container's
axis with the side wall. Unlike the first container, which has uncountably many
vertical planes of symmetry going through the container's axis, the second one has
only one plane of symmetry in which the baffle lies. The effect of broken symmetry
leads to the essential difference of properties of sloshing eigenvalues and
eigenfunctions; the most important of which are the following.

First, all eigenvalues of the container with baffle are simple, whereas in the
absence of baffle each eigenvalue has multiplicity two except for those corresponding to axisymmetric eigenmodes. This fact has essential
influence on properties of eigenfunctions discussed below. Second, it occurs that
the lowest eigenvalue is substantially smaller in the presence of baffle comparing
with the case when there is no baffle. Third, the set of eigenvalues of the
container without baffle is a subset of the set existing when the baffle is present;
the difference between the latter and former sets is an infinite set. The elements
of both subsets are intermittent without any apparent pattern.

Comparing properties of eigenfunctions, we see that of two linearly independent
eigenfunctions, corresponding to every eigenvalue in the case when there is no
baffle, one is an even function of $y$, whereas the other is odd. Therefore,
locations of maxima and minima of the corresponding free surface elevation and of
its nodes can be chosen arbitrarily. This follows from the fact that there are
uncountably many linear combinations of linearly independent eigenfunctions. 

On the other hand, every eigenfunction, existing in the presence of baffle, is
either odd or even function of $y$ provided the baffle lies in the $(x, z)$-plane.
This leads to the completely different behaviour of sloshing modes in this case. For
example, the elevation of the free surface, corresponding to the fundamental
sloshing mode, has its maximum and minimum attained at the points, where the baffle
is attached to the side wall. This, along with the diminished lowest eigenvalue, is
the most important effect that results from the symmetry breaking by the radial
baffle.

In $\S$~4, similar results were obtained for an annular container with and without baffle. 
It is natural that behaviour of eigenvalues and eigenfunctions is more complicated, in particular including dependence on an additional parameter of the problem (thickness of the annulus). However, observations of effects of broken symmetry\,---\,as the baffle is added\,---\,are very similar to the case of a circular container. 

%
%

\end{document}